\documentclass[conference]{IEEEtran}

\ifCLASSINFOpdf
   \usepackage[pdftex]{graphicx}
 \else

\fi

\usepackage{array}
\usepackage{subfigure}
\usepackage{amsmath,amsthm}

\usepackage{amstext,amssymb}

\newtheorem{proposition}{Proposition}

\theoremstyle{definition}
\newtheorem{definition}{Definition}

\newtheorem{example}{Example}

\begin{document}

\title{Correct Undetected Errors with List Decoding in ARQ Error-control Systems}

\author{\IEEEauthorblockN{Jingzhao Wang}
\IEEEauthorblockA{Department of Computer Science and\\
Engineering\\
Shanghai Jiao Tong University\\
Shanghai 200240, China\\
Email: wangzhe.90@sjtu.edu.cn}
\and
\IEEEauthorblockN{Yuan Luo}
\IEEEauthorblockA{Department of Computer Science and\\
Engineering\\
Shanghai Jiao Tong University\\
Shanghai 200240, China\\
Email: yuanluo@sjtu.edu.cn}
}

\maketitle

\begin{abstract}
Undetected errors are important for linear codes, which are the only type of errors after hard decision and automatic-repeat-request (ARQ), but do not receive much attention on their correction. In concatenated channel coding, sub-optimal source coding and joint source-channel coding, constrains among successive codewords may be utilized to improve decoding performance. In this paper, list decoding is used to correct the undetected errors. The benefit proportion of the correction is obviously improved especially on Hamming codes and Reed-Muller codes, which achieves about $40\%$ in some cases. But this improvement is significant only after the selection of final codewords from the lists based on the constrains among the successive transmitted codewords. The selection algorithm is investigated here to complete the list decoding program in the application of Markov context model. The performance of the algorithm is analysed and a lower bound of the correctly selected probability is derived to determine the proper context length.

\end{abstract}

\section{Introduction}
\label{intro}
Error detection using block codes combined with repeat requests is a widely used method of error control in communication systems\cite{Klove:2007:CED:1526180}, which is called automatic-repeat-request (ARQ) scheme.\par Undetected errors are the only type of errors after hard decision and automatic-repeat-request (ARQ), where the sent codeword is transmitted into another codeword.
List decoding was introduced independently by Elias \cite{Elias57} and Wozencraft \cite{Woz58} in the late 1950's. For a received vector, the decoder doesn't output a single codeword, but outputs a list of possible codewords. It is considered to be list decoded correctly if the sent codeword is in the list. This paper uses list decoding to correct undetected errors of ARQ systems and give a list of possible codewords for each received codeword after retransmissions. \par
In some cases, there may be constraints among successive codewords, i.e. the input codewords are dependent. For example, when concatenated codes \cite{forney1966} are used in channel coding, the codewords of the inner code must be constrained by the outer code. In joint source-channel coding, there also be memory structure inherent with in the sequence output from the source \cite{JointSC}. Moreover, in the model for storage or communication, many files are not compressed or compressed sub-optimally in source coding \cite{jiang2015NMVW},\cite{jiang2015ITW},\cite{weissman2005}. In this situation, there may also be constraints among the successive transmitted codewords , where redundancy mining can be used for error correction, which is considered in \cite{jiang2015NMVW},\cite{jiang2015ITW}. Inspired by these works, we assume the code context model is a Markov chain which is an elementary stochastic model used in natural language processing models, and investigate an algorithm to select codewords from the lists.

This paper is organized as follows. Section \ref{sec:2} presents some necessary preliminaries and notations. In Section \ref{sec:3} the error correcting probability by list decoding of linear codes is analysed. And Hamming codes and Reed-Muller codes are investigated as numerical examples where the benefit proportion of list decoding can achieve about $40\%$ in some cases. In Section \ref{sec:4}, an algorithm based on context is investigated to complete list decoding program, with which the most probable codewords are selected from the lists. The algorithm performance is analysed in Section \ref{sec:5}.  Finally, we conclude this paper in Section \ref{Con}.
\section{Preliminaries and Notations}
\label{sec:2}
The communication channel used in this paper is binary symmetric channel (BSC) with crossover probability $p$. The encoder is a one-to-one map function from the message set $\mathcal{M}=\{0,1,...,M-1\}$ to the codebook $C$, where $C$ is a subset of $\{0,1\}^n$ and $n$ is known as the blocklength of code $C$ .

$w_H(\mathbf{x})$ is the Hamming weight of a vector $\mathbf{x}$ and $d_H(\mathbf{x},\mathbf{y})$ is the Hamming distance between two vectors $\mathbf{x},\mathbf{y}\in \{0,1\}^n$.

For a binary $[n,k,d]$ linear code $C$ with minimum distance $d=\min\{w_H(\mathbf{x})|\mathbf{x}\in C,\mathbf{x}\neq \mathbf{0}\}$, the weight distribution is defined as
\begin{align*}
A_i^w=A_i^w(C)=\#\{\mathbf{x}\in C|w_H(\mathbf{x})=i\}
\end{align*}
and the weight distribution function is defined as
\begin{align*}
A^w_C(x,y)=\sum_{i=0}^nA^w_ix^{n-i}y^i.
\end{align*}\par
%

\section{The error probability of ARQ systems with list decoding}
\label{sec:3}
In this section, we first analyse the error correcting probability by list decoding of linear codes, which can be calculated by Proposition \ref{listr}. Then the error probability of ARQ systems is calculated in Proposition \ref{Pb}. In subsection \ref{subsec:3.1} and \ref{subsec:3.2}, the benefit proportion of error probability after list decoding is illustrated in Fig. \ref{Hamming} and Fig. \ref{RMl}, where Hamming codes and Reed-Muller codes are investigated as numerical examples.\par

Assume $\mathbf{X}=(X_1,X_2,...,X_n) \in \{0,1\}^n$ is sent and $\mathbf{Y}=(Y_1,Y_2,...,Y_n)\in \{0,1\}^n$ is received, where $\mathbf{X}$ and $\mathbf{Y}$ are vectors of random variables. Let $B(\mathbf{y},r)=\{\mathbf{z}\in \{0,1\}^n:\mathbf{y}\in \{0,1\}^n, d_H(\mathbf{z},\mathbf{y})\leq r\}$ denote the sphere of radius $r$ about $\mathbf{y}$ and consider the conditional probability as follows:
\begin{align}
&\text{Pr}\{\mathbf{Y}\in C\ \text{and}\  \mathbf{Y}\neq 0^n, 0^n \in B(\mathbf{Y},r)|\mathbf{X}=0^n\}\nonumber\\
&=\text{Pr}\{\mathbf{Y}\in C\ \text{and}\  \mathbf{Y}\neq 0^n, \mathbf{Y}\in B(0^n,r)|\mathbf{X}=0^n\},\label{ytoc}
\end{align}
where $r\geq d$. $0^n \in B(\mathbf{Y},r)$ and $\mathbf{Y}\neq 0^n$ mean that an undetected error occurs when  $\mathbf{X}=0^n$ is sent and the correct codeword $0^n$ is in the list with decoding radius $r\ (r\geq d)$. In this case, it is defined to be list decoded correctly.\par
For all the codewords in $C$, the probability in (\ref{ytoc}) can be described as follows:
\begin{align}
P^{List}(r)=\sum_{\mathbf{c}\in C,\mathbf{c}\neq 0^n}\text{Pr}\{\mathbf{Y}=\mathbf{c}, \mathbf{c}\in B(0^n,r)|\mathbf{X}=0^n\}, \label{Ps}
\end{align}
where $r\geq d$. $P^{List}(r)$ is called the error correcting probability by list decoding, which is also the reduced probability of undetected error, see Proposition \ref{listr}.

\noindent {\bf Remark.} As is shown in \cite{guruswami2001list}, the list size should not be too large and be at most a polynomial in the blocklength. So this paper considers the cases that the decoding radius $r$ equals to d, the minimum distance of code $C$. Note that, the completing of list decoding program will be provided in Section \ref{sec:4} by using context.

\begin{proposition}\label{listr}
Assume the channel is BSC with crossover probability $p$. Then, the error correcting probability by list decoding with radius $r=d$ is
\begin{align*}
P^{List}&\triangleq P^{List}(d)\\
&=\sum_{\mathbf{c}\in C,\mathbf{c}\neq 0^n}Pr\{\mathbf{Y}=\mathbf{c}, \mathbf{c} \in B(0^n,d)|\mathbf{X}=0^n\} \nonumber\\
&=A_d^w\cdot p^d\cdot (1-p)^{n-d}.
\end{align*}
The list size is $L=A_d^w+1$.
\end{proposition}

For ordinary linear codes, the computation of $A_i^w(0\leq i\leq n)$ is an NP-hard problem. But for some special codes, the weight distributions are known and can be used to calculate $P^{List}$.\par
In ARQ systems, the error probability is defined in \cite{Lin:2004:ECC:983680}, see
\begin{equation}
P_e^{ARQ}=\frac{P_{ue}}{P_{ue}+P_c},\label{Pe}
\end{equation}
where $P_{c}$ denotes the probability that a received vector contains no error and $P_{ue}$ denotes the probability that a received vector contains an undetected error pattern.\par
After list decoding in ARQ system, the $P_{ue}$ is reduced to $P_{ue}^{List}=P_{ue}-P^{List}$ and the error probability changes from (\ref{Pe}) to
\begin{align}
P_e^{ARQ,List}=\frac{P_{ue}^{List}}{P_{ue}+P_c}.\label{Puee}
\end{align}
The benefit proportion is defined by
\begin{align}
P_b=\frac{P_e^{ARQ}-P_e^{ARQ,List}}{P_e^{ARQ}}.\label{Pb1}
\end{align}
The results (\ref{Pe}), (\ref{Puee}) and (\ref{Pb1}) form the following proposition.
\begin{proposition}\label{Pb}
If the crossover probability of BSC is $p$, it is clear that $P_c=(1-p)^n$, $
P_{ue}=A^w_C(1-p,p)-P_c=A^w_C(1-p,p)-(1-p)^n$. The probability of undetected error after list decoding is $P_{ue}^{List}=P_{ue}-P^{List}$. After ARQ retransmissions, assume all the received vectors are codewords, then the error probability is
\begin{align*}
P_e^{ARQ}&=\frac{P_{ue}}{P_{ue}+P_c}\nonumber\\
   &=\frac{A^w_C(1-p,p)-(1-p)^n}{A^w_C(1-p,p)},
\end{align*}
the error probability after list decoding is
\begin{align*}
P_e^{ARQ,List}&=\frac{P_{ue}^{List}}{P_{ue}+P_c}\nonumber\\
         &=\frac{A^w_C(1-p,p)-(1-p)^n-A_d^w\cdot p^d\cdot (1-p)^{n-d}}{A^w_C(1-p,p)},
\end{align*}
and the benefit proportion is
\begin{align*}
P_b&=\frac{P_e^{ARQ}-P_e^{ARQ,List}}{P_e^{ARQ}}\nonumber\\
   &=\frac{A_d^w\cdot p^d\cdot (1-p)^{n-d}}{A^w_C(1-p,p)-(1-p)^n}.
\end{align*}
\end{proposition}
Two numerical examples of Proportion \ref{Pb} in ARQ systems are given in Subsection \ref{subsec:3.1} and \ref{subsec:3.2}, respectively.
\subsection{Hamming Codes}
\label{subsec:3.1}
Next we illustrate the benefits of list decoding on the error probability $P^{ARQ}_e$ of $[n=2^m-1,n-m]$ Hamming code. The weight distribution function of Hamming code is known \cite{MW} and the corresponding list size in Proposition \ref{listr} is $L=A_3^w+1$.

\begin{figure}[htbp]
  \centering
  \includegraphics[width=0.24\textwidth]{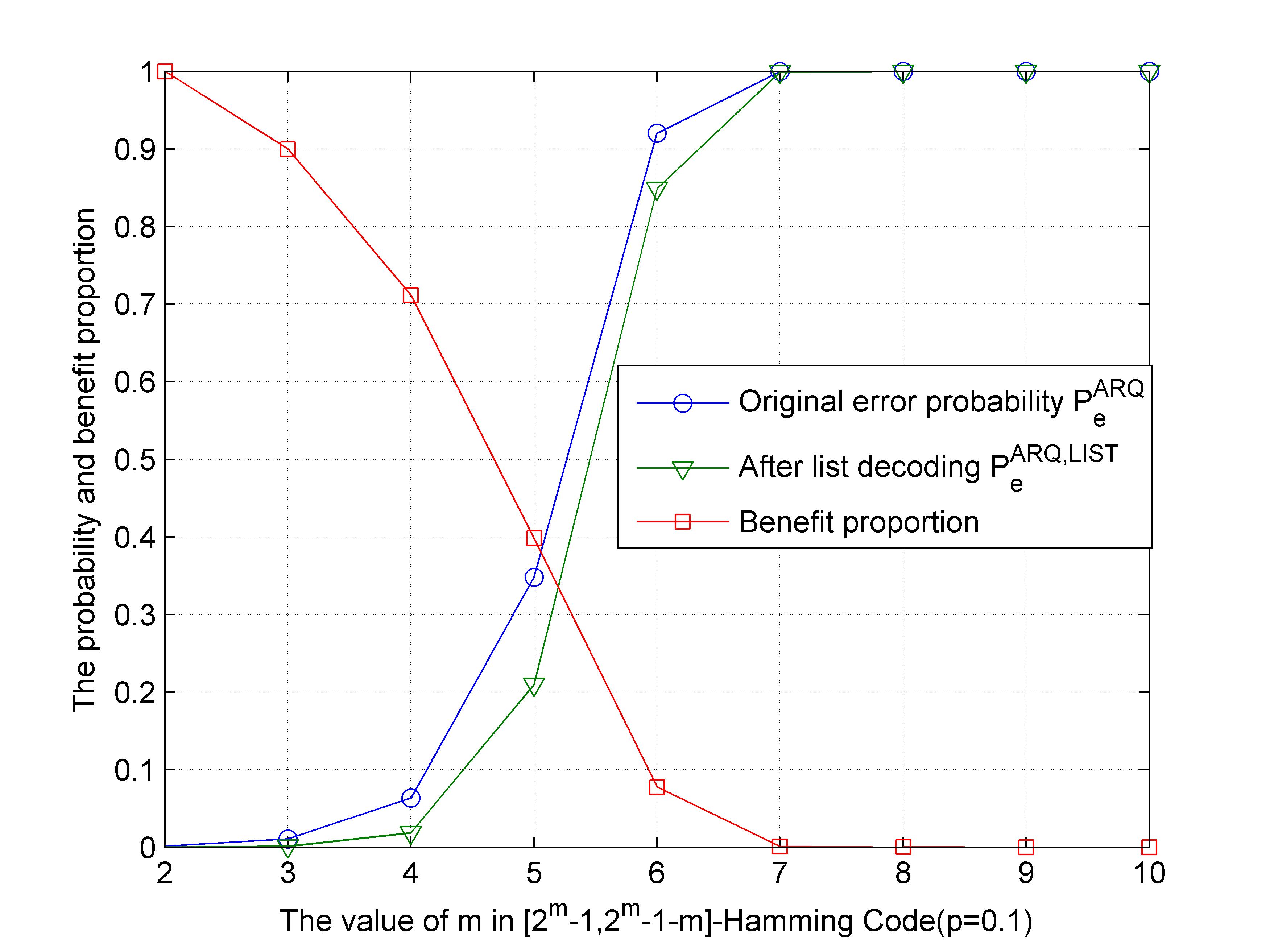}
  \includegraphics[width=0.24\textwidth]{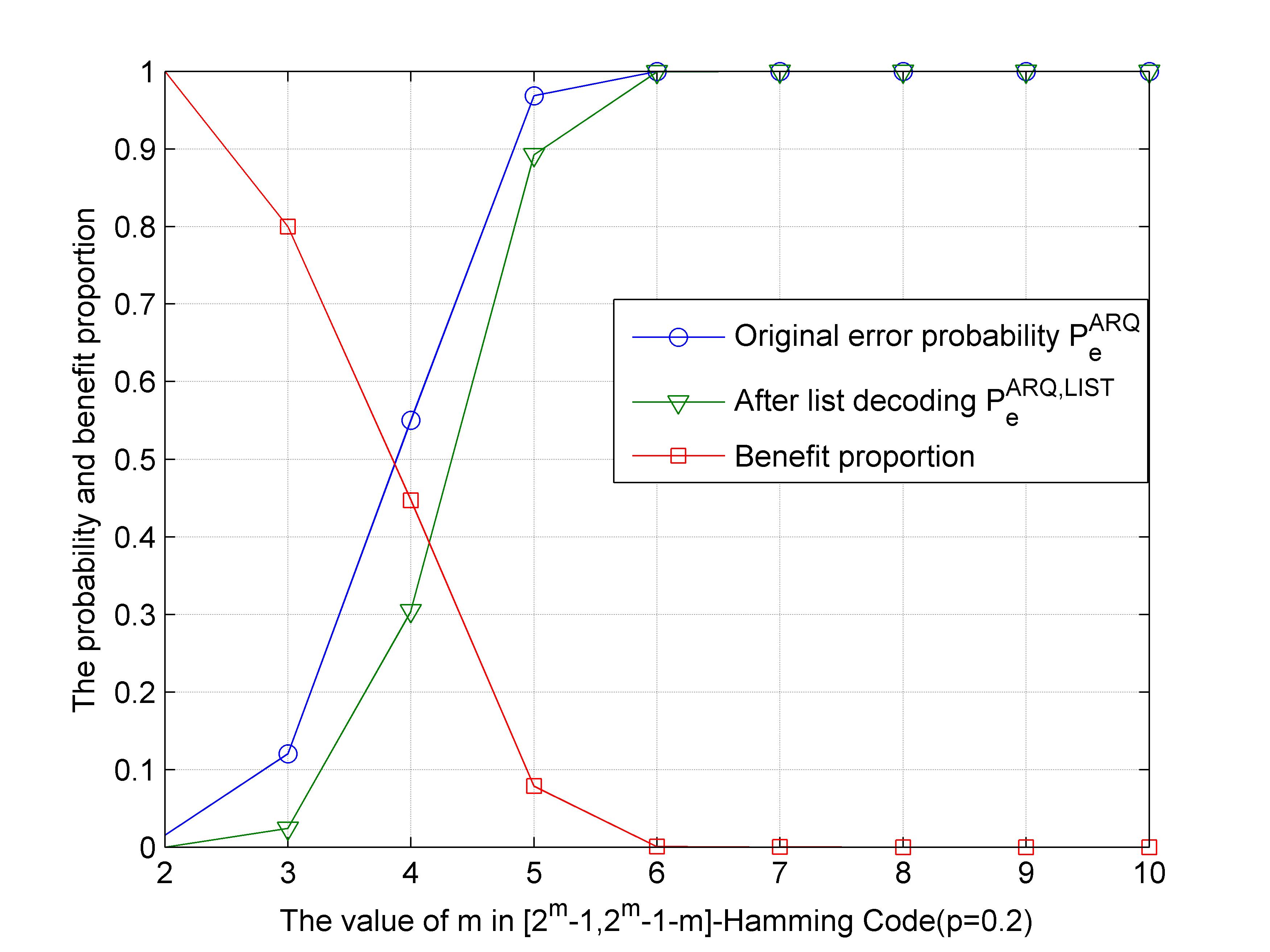}
  \caption{The Numerical Results of Hamming Codes}\label{Hamming}
\end{figure}

The parameters of Proposition \ref{Pb}, i.e. $P_e^{ARQ}$, $P_e^{ARQ,List}$ and $P_b$, are illustrated in Fig. \ref{Hamming}, where the messages are transmitted through BSC with crossover probability $p$.
As is shown by the figure, $P_e^{ARQ}$ increases with block length. For Hamming codes of $m=4,5,6\ (p=0.1)$ or $m=3,4,5\ (p=0.2)$, the benefit of list decoding is obvious. Especially for $m=5\ (p=0.1)$ and $m=4\ (p=0.2)$, $P_e^{ARQ}$ is large and the benefit proportion $P_b$ is about $40\%$, a meaningful amount.
\subsection{Reed-Muller Codes}
\label{subsec:3.2}
As is shown in \cite{MW}, the weight distribution of RM codes is partly known. In most cases, the higher the code rate is, the higher the undetected error probability is, and correcting undetected errors is significant when the probability is high. Let $RM(r,m)$ denote the $r^{th}$ order binary Reed-Muller code with block length $n=2^m$.

\begin{figure*}[!t]
\centering
\subfigure[RM(2,4)]
{\includegraphics[width=2.3in]{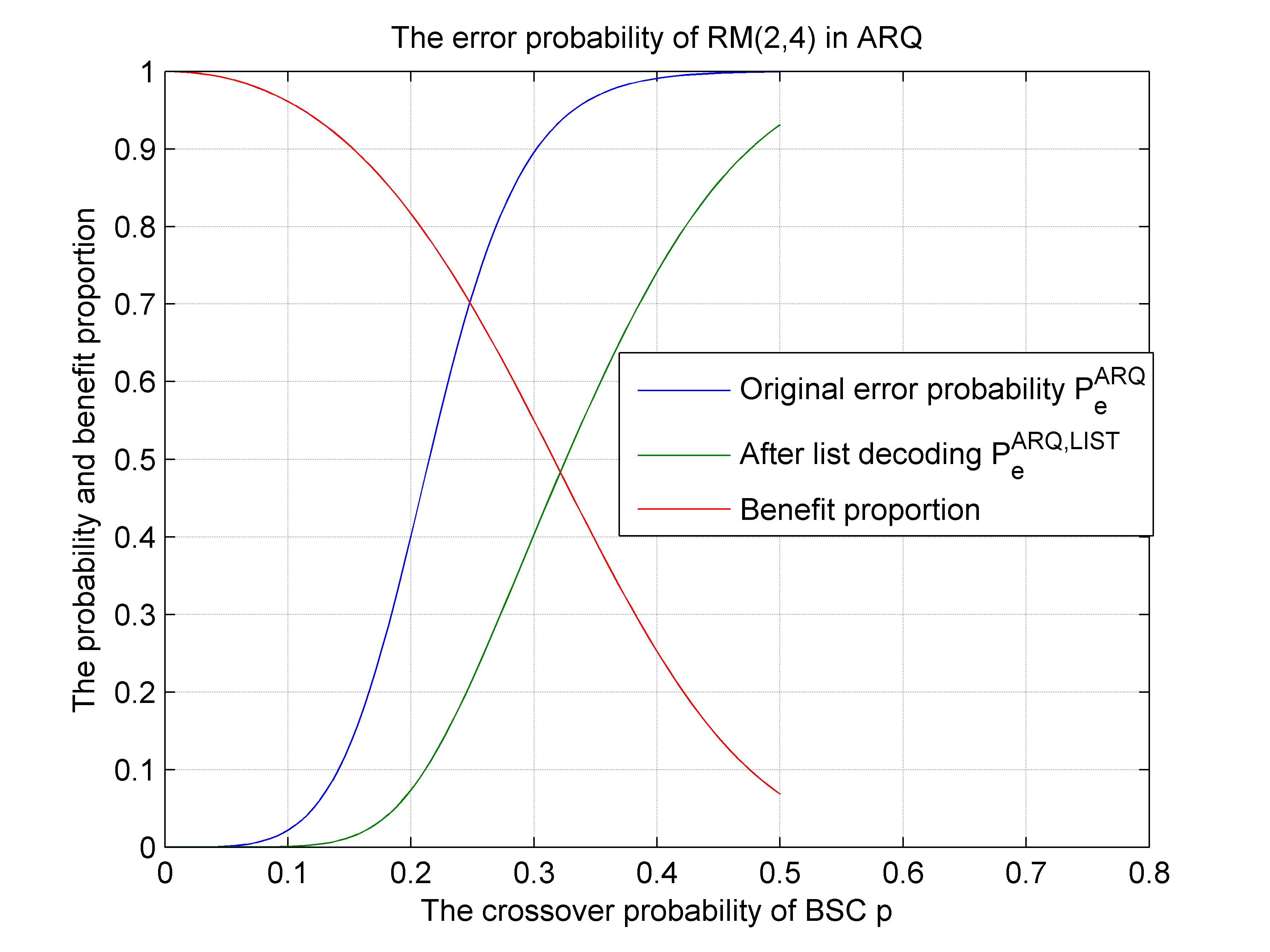}}
\subfigure[RM(2,5)]
 {\includegraphics[width=2.3in]{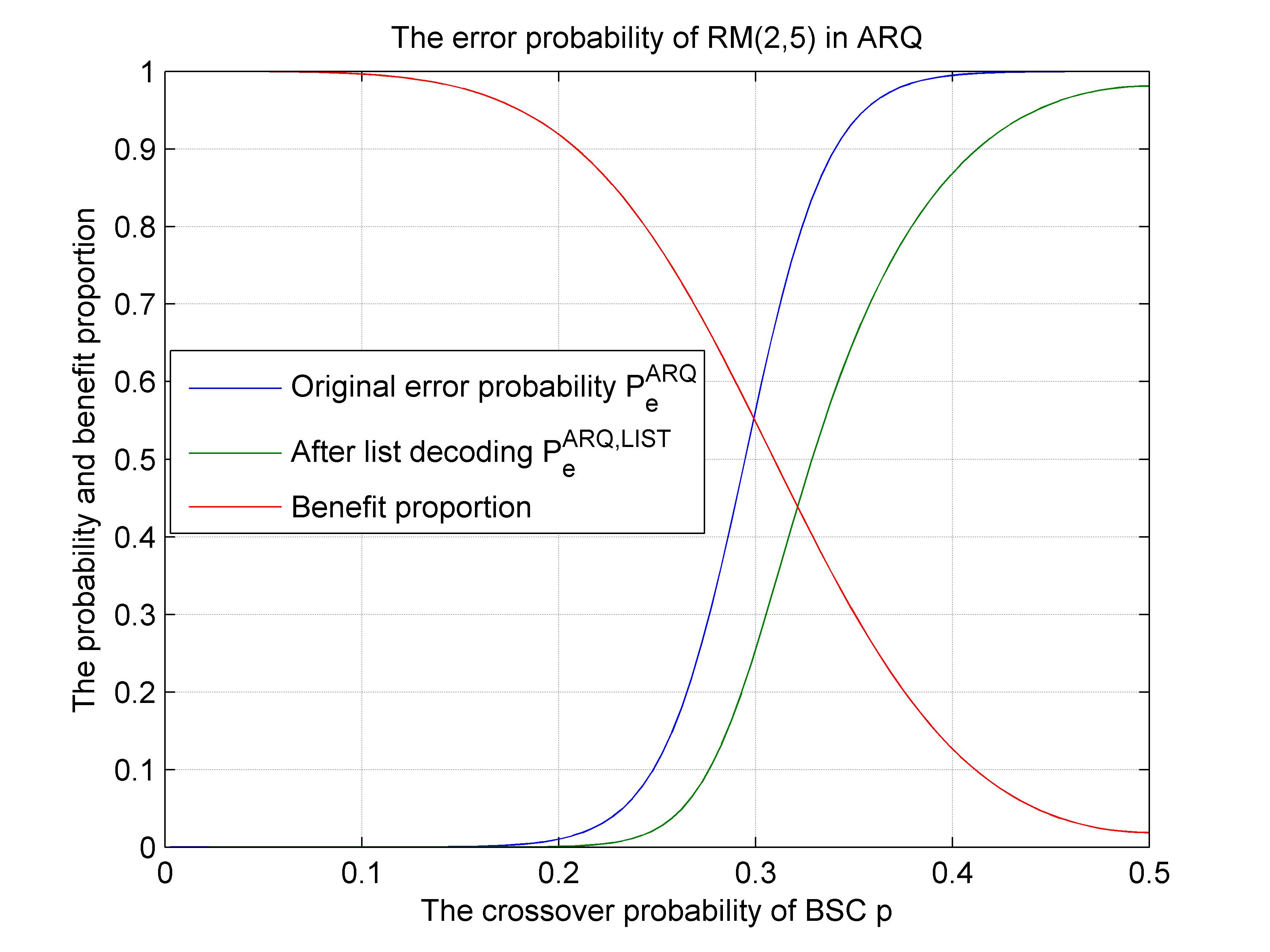}}
\subfigure[RM(3,5)]
{\includegraphics[width=2.3in]{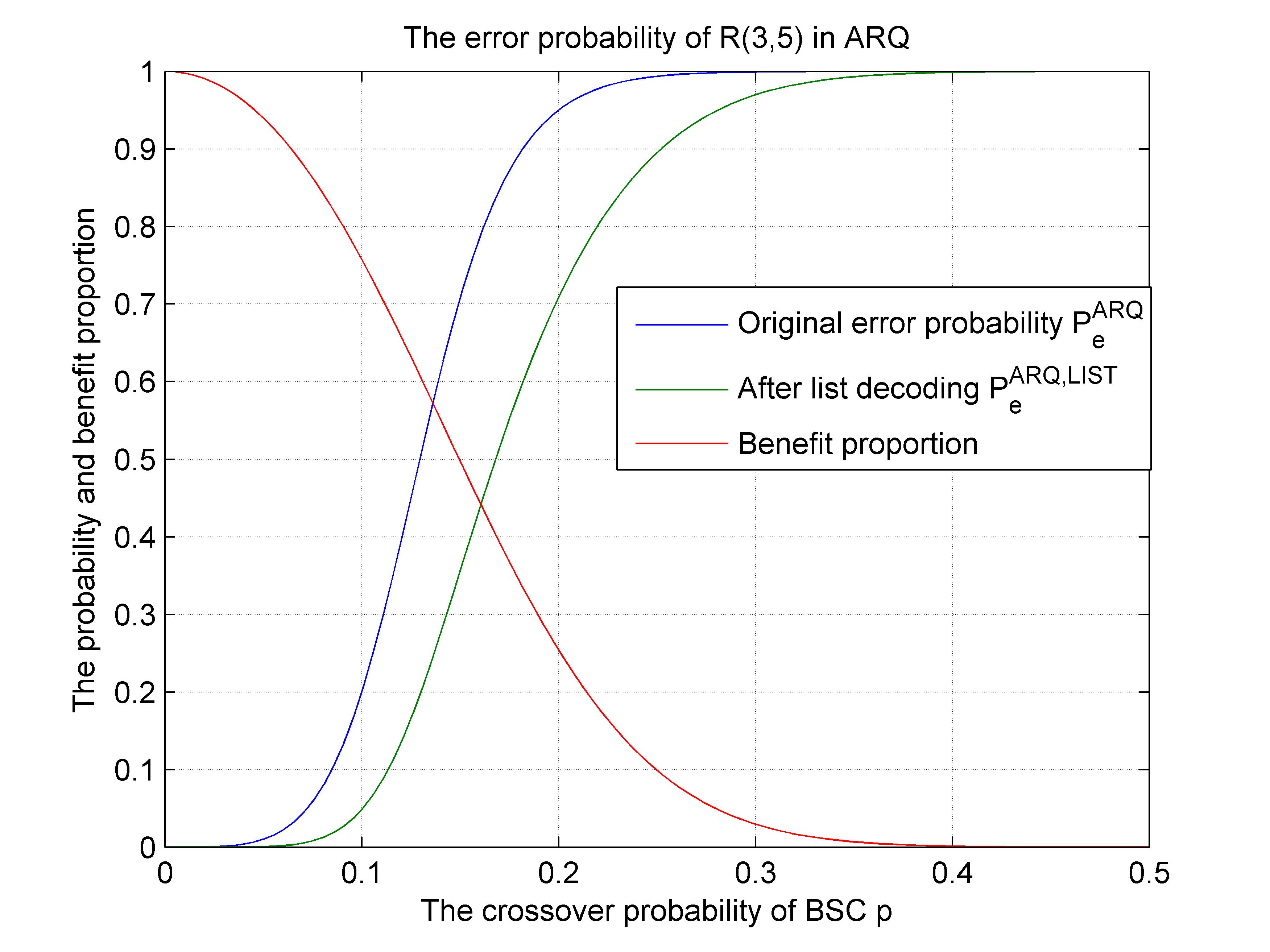}}
\centering
\subfigure[RM(4,6)]
{\includegraphics[width=2.3in]{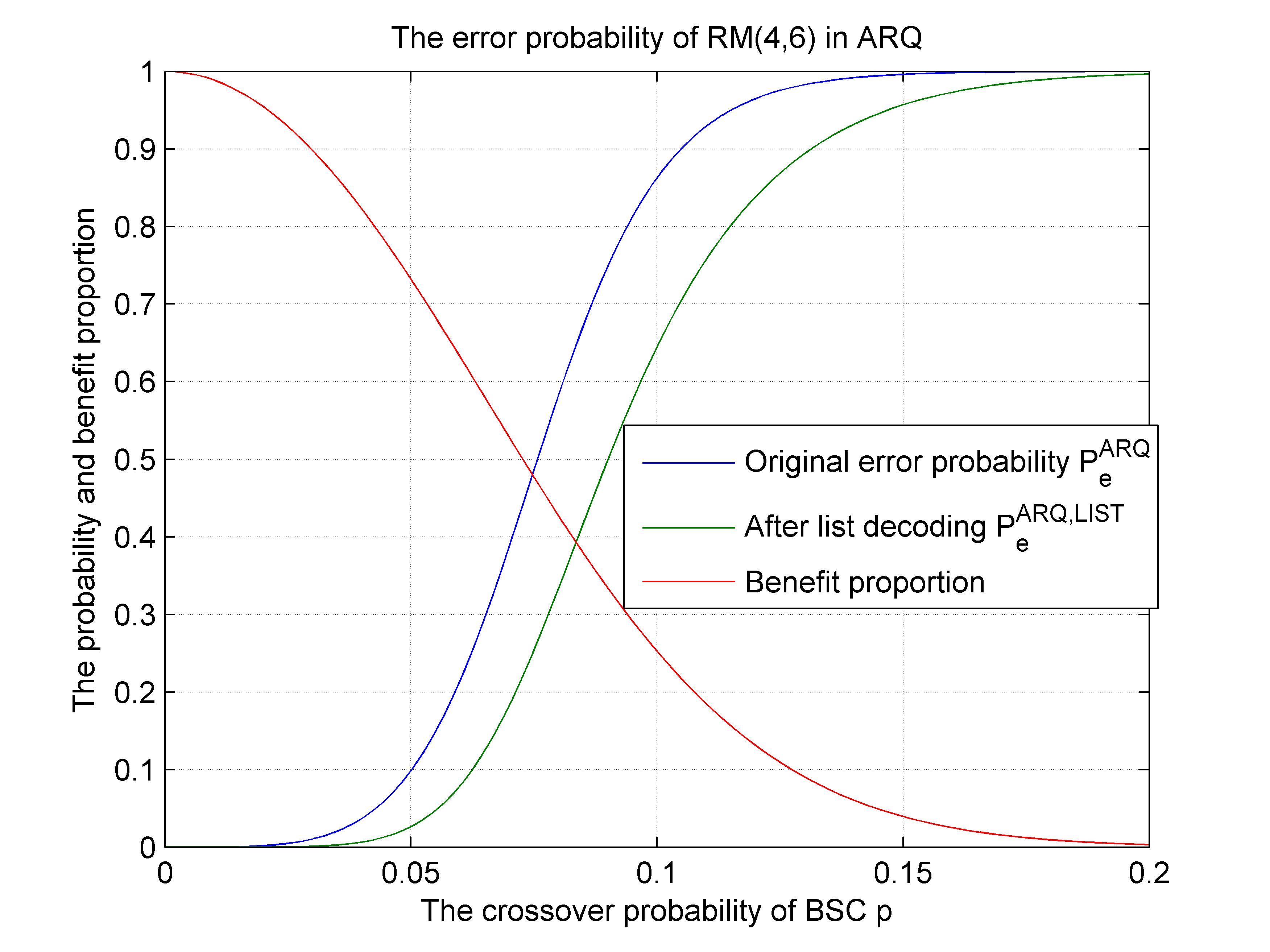}}
\subfigure[RM(5,7)]
 {\includegraphics[width=2.3in]{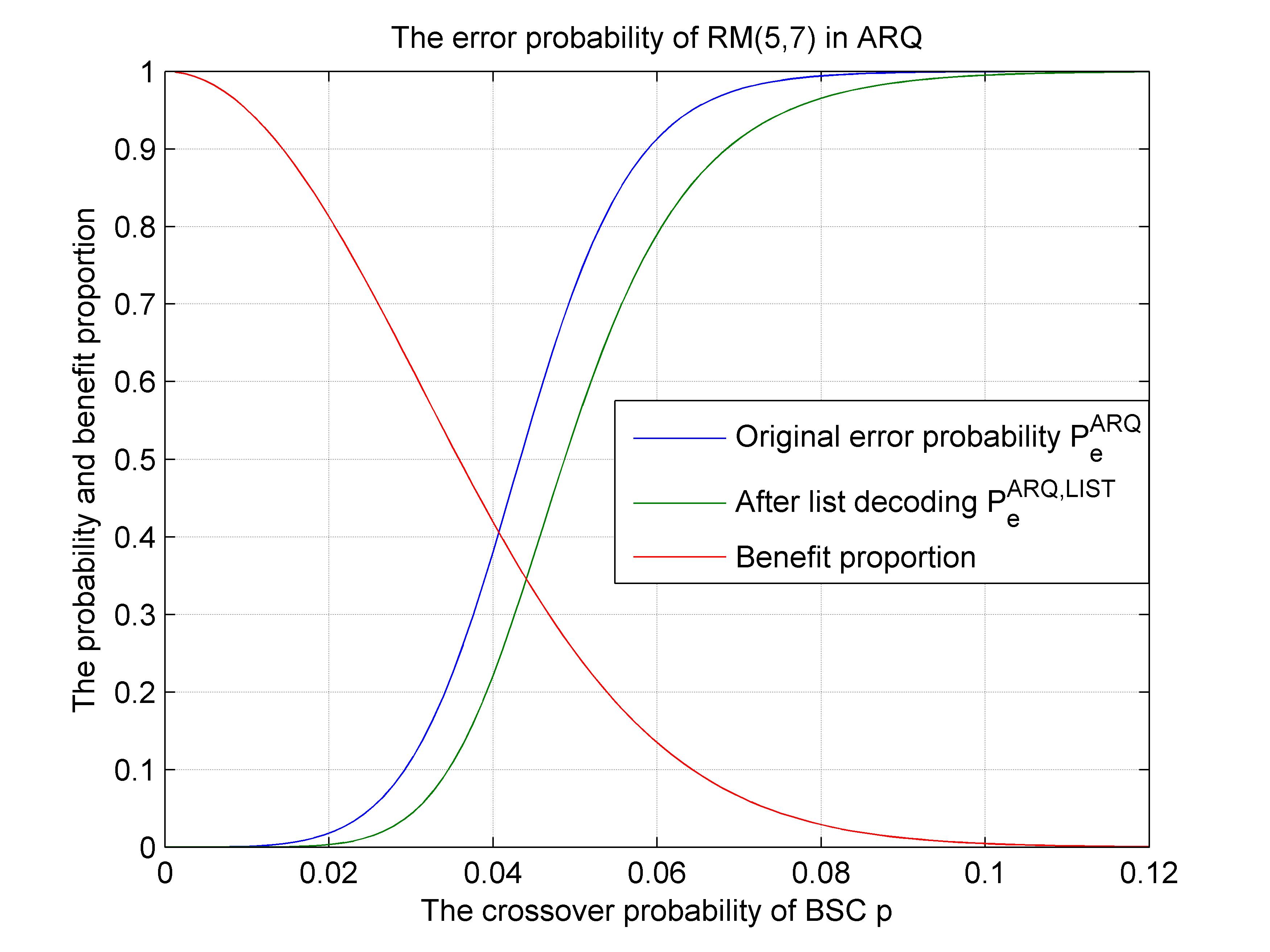}}
\subfigure[RM(6,8)]
{\includegraphics[width=2.3in]{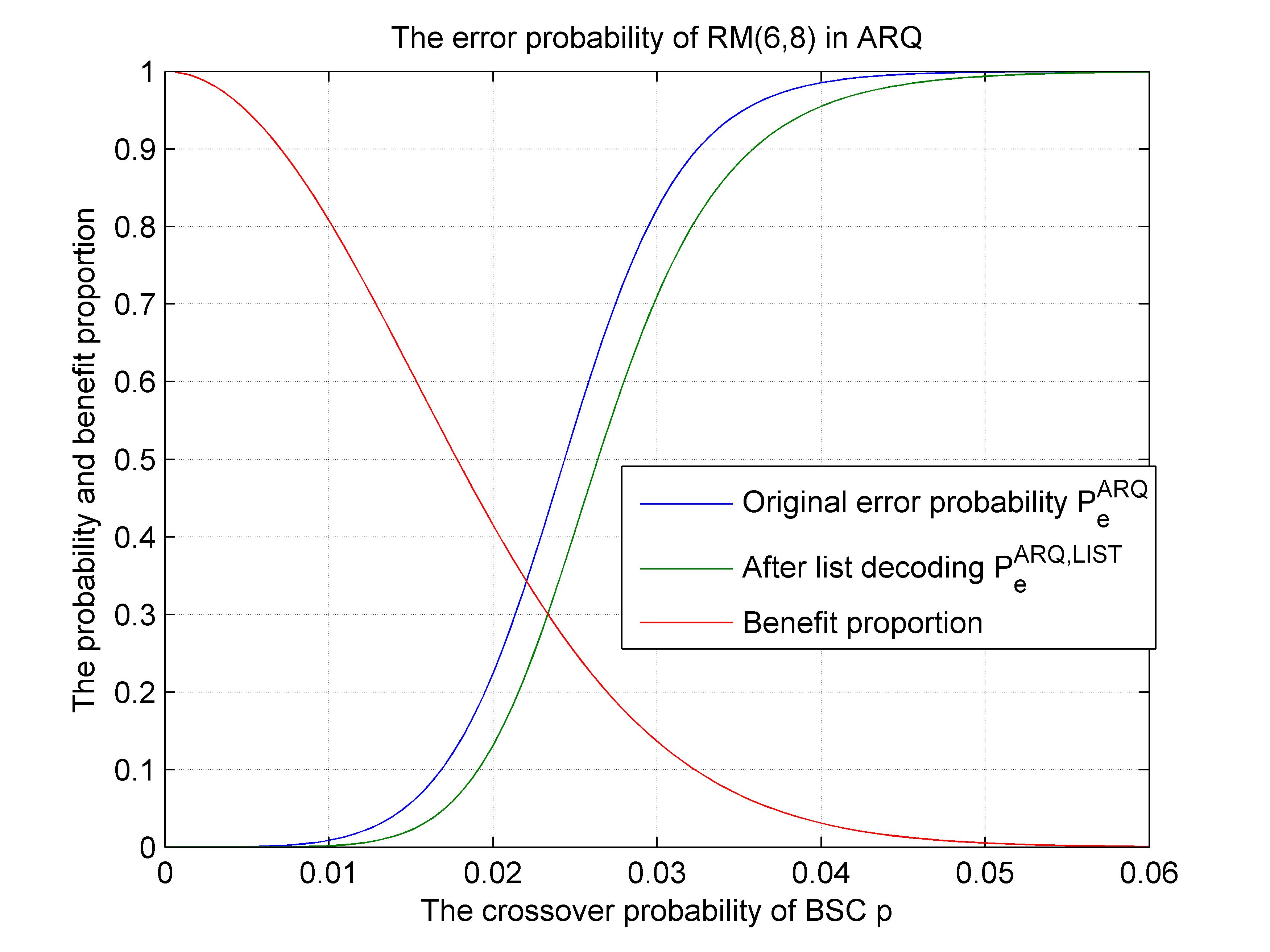}}
\caption{ The Numerical Results of Reed-Muller Codes}\label{RMl}
\end{figure*}

The numerical results of $P_e^{ARQ}$, $P_e^{ARQ,List}$ and $P_b$ in Proposition \ref{Pb} are illustrated in Fig. \ref{RMl}, where we can see that the benefit proportion is decreasing with the crossover probability $p$. And for Reed-Muller codes of high rate, the benefit of list decoding is evident for small $p$, which is supported by Fig. \ref{RMl} (a) and (d)-(f). When $P_e^{ARQ}$ is less than $0.4$, the benefit proportion is more than $40\%$ shown in Fig. \ref{RMl} (a)-(e).

\section{List-decoding Based on Context}
\label{sec:4}
In this section, an algorithm based on the context is investigated and dubbed \emph{context list decoding algorithm (CLDA)}. The code context is described with Markov chain satisfying equation (\ref{Markov}). The performance of CLDA depends on the context structure. The process of CLDA is shown in Algorithm 1 and Example 1, with which the most probable codewords are selected from the lists.

The appearance of a codeword may be affected by former or latter codewords, which is called the \emph{context} of a code. For instance, the appearance of each English letter in a word depends on its former and latter letters. At the same time, an English word is also affected by its adjacent words in a sentence. In natural language processing models, Markov chain is primary. Therefore, we assume the code context model is Markov.

Then the context of all the received codewords can be seen as a stochastic process $\{X_t,\ t=0,1,2,...\}$ that takes values on $\mathcal{M}=\{0,1,...,M-1\}$. Each codeword is bijectively mapped to a state of the process. After retransmissions in ARQ systems, all the received vectors are codewords. So in the following part, we use the state in $\mathcal{M}$ to represent the corresponding codeword and give a list of all possible codewords for each received codeword. The proper ones are selected from the lists based on the context. The selection of codewords from lists is equivalent to the determination of the states, see the following part.

Assume the process is a Markov chain and let $P_{ij}$ denote the one-step transition probability from state $i$ to state $j$, then
\begin{align}
P_{ij}=\text{Pr}\{X_{t+1}=j|X_t=i\}, \label{Markov}
\end{align}
for all states $i, j \in \mathcal{M}$ and all $t\geq 0$.
Let $\mathbf{P}$ denote the the matrix of $P_{ij}$, then
\begin{align*}
\mathbf{P}=(P_{ij})_{M\times M}(0\leq i,j\leq M-1).
\end{align*}


The context of a code can be represented by $\mathbf{P}$, with which Algorithm 1 (CLDA) will select the most probable queue of codewords. To describe CLDA, some definitions are given as follows.
\begin{definition} \label{def_sentence}
A queue of successive codewords with context constraint is called a \emph{codeword sentence}. If the queue length is $N$, it is said to be an $N$-\emph{codeword sentence}. The appearance probability of a codeword sentence is dubbed \emph{codeword sentence weight}.
\end{definition}
Then, the weight of $N$-\emph{codeword sentence} $(i_0,i_1,...,i_{N-1})$ is
\begin{align}
&\text{Pr}\{X_0=i_0,...,X_{N-1}=i_{N-1}\}\nonumber\\
&=\Pr\{i_0\}\cdot\Pr\{i_{1}|i_{0}\}\cdot...\cdot\Pr\{i_{N-2}|i_{N-1}\}\notag\\
&=\pi_{i_0}\cdot P_{i_{1},i_{0}}\cdot...\cdot P_{i_{N-2},i_{N-1}},\label{equ_Mark}
\end{align}
where let $\text{Pr}\{i_0\}\triangleq\text{Pr}\{X_0=i_0\}\triangleq\pi_{i_0}$ denote the priori probability of $i_0$ and let $\text{Pr}\{i_{k+1}|i_k\}\triangleq\text{Pr}\{X_{k+1}=i_{k+1}|X_{k}=i_k\}$.

For linear codes, every given list contains the same number of codewords and $L$ is the list size. Let $\mathbf{r}=(r_0,r_1,...,r_{N-1})$ denote the received codeword sentence and $N$ is the codeword sentence length. Let $\mathbf{l}=(l_0,l_1,...,l_{N-1})$ denote the codeword sentence selected from the successive lists, then the problem of finding the most probable $N$-codeword sentence in the lists is equivalent to compute
\begin{align*}
 \mathop{\arg\max}_{(l_0,l_1,...,l_{N-1})\in L_0\times L_1\times ...\times L_{N-1}} \text{Pr}\{(l_0,l_1,...,l_{N-1})\},
\end{align*}
where $L_0,L_1,...,L_{N-1}$ are successive lists given by $\mathbf{r}$ as follows:
\begin{align}
&L_i=\mathcal{L}(r_i),0\leq i\leq N-1,\notag\\
&\mathcal{L}(r_i)=\{c\in C: d_H(c,r_i)\leq d\}\label{equ_L}.
\end{align}

Let function $T(l_i)$ compute the maximal codeword sentence weight when some codeword $l_i$ is selected form List $L_i$ and appended to the previously selected codeword sentence $(l_0,l_1,...,l_{i-1})$. For $i\in[2,N-1]$, we have
\begin{align*}
T(l_i)\triangleq \max_{(l_0,l_1,...,l_{i-1})\in L_0\times L_1\times...\times L_{i-1}}\text{Pr}\{(l_0,l_1,...,l_i)\}.
\end{align*}
Then
\begin{small}
\begin{align*}
&T(l_i)=\max_{\substack{(l_0,l_1,...,l_{i-1})\in\\ L_0\times L_1\times...\times L_{i-1}}}\text{Pr}\{(l_0,l_1,...,l_{i-1})\}\text{Pr}\{l_{i}|l_{i-1}\}\\
&=\max_{l_{i-1}\in L_{i-1}}\Big(\text{Pr}\{l_i|l_{i-1}\}\cdot
\max_{\substack{(l_0,l_1,...,l_{i-2})\in\\ L_0\times L_1\times...\times L_{i-2}}}\text{Pr}\{(l_0,l_1,...,l_{i-1})\}\Big)\\
&=\max_{l_{i-1}\in L_{i-1}}\text{Pr}\{l_i|l_{i-1}\}T(l_{i-1})\label{equ_T}
\end{align*}\end{small}
and $T(l_0)=\text{Pr}\{l_0\}=\pi_{l_0}$.

The derived recurrence suggests that the optimization problem can be solved with dynamic programming algorithm specified in Algorithm 1 (CLDA), which is adapted from the Viterbi decoding \cite{Viterbi1967}. The final solution is computed iteratively, starting from $T(l_1)$ according to the recurrence. When the last iteration is finished, we trace back along the path with the maximal codeword sentence weight, selecting the most probable codeword sentence.


Let $\mathbf{L}=(l_{ij})_{L\times N}(0\leq i\leq L-1,0\leq j\leq N-1)$ store the values of each node. Let $\mathbf{P}=(P_{ij})_{M\times M}(0\leq i, j\leq M-1)$ denote the codeword transition probability matrix and $\Pi=(\pi_0,\pi_1,...,\pi_{M-1})^T$ denote the priori probability of each codeword, which are decided by the context of the codewords. The CLDA is as follows:\\

\noindent\textbf{Algorithm 1(CLDA).}\label{algo_1}\par
In the following four steps, first, $\mathbf{P}$, $\mathbf{L}$ and $\Pi$ defined above can be used to calculate a weight matrix $\mathbf{T}$ and a jumping point matrix $\mathbf{D}$. Then, matrix $\mathbf{R}$, which stores the jumping points of the path with length N and the largest weight, is determined by $\mathbf{T}$ and $\mathbf{D}$. At last, the final codewords after context list decoding are stored in $\mathbf{F}$ calculated by $\mathbf{R}$ and $\mathbf{L}$.

1. Initialize matrix $\mathbf{T}=(t_{ij})_{L\times N}$ where $t_{ij}$ stores the maximum codeword sentence weight from the nodes in List $0$ to the $i$th node in the current List $j$. The first column of $\mathbf{T}$ is initialized with $(\pi_{l_{0,0}},\pi_{l_{1,0}},...,\pi_{l_{L-1,0}})^T$, namely, $t_{i,0}=\pi_{l_{i,0}}(0\leq i\leq L-1)$ and others with zeroes.\par
2. From List $0$ to List $N-1$, compute the codeword sentence weight:
\begin{equation}
t_{ij}=\max\limits_{0\leq k\leq L-1} (t_{k,j-1}\cdot P_{l_{k,j-1},l_{ij}}),\label{equ_t}
\end{equation}
where $0\leq i\leq L-1,1\leq j\leq N-1$.\par
The corresponding $k$ which achieves the maximum value is stored in matrix $\mathbf{D}=(x _{ij})_{L\times N}$:
\begin{equation}
x_{ij}=\arg\max\limits_{0\leq k\leq L-1} (t_{k,j-1}\cdot P_{l_{k,j-1},l_{ij}}),\label{equ_x}
\end{equation}
where $0\leq i\leq L-1,1\leq j\leq N-1$.\par
\noindent {\bf Remark.}
\begin{itemize}
\item $k$ is the row (node) index and $j$ is the column (list) index;
\item $l_{ij}$ is the value (codeword) of the $i$th node in List $j$;
\item $x_{ij}$ is the jumping point from List $j-1$ to the $i$th node in List $j$, which reaches the largest weight to current node.
\item we select one codeword randomly if there are more than one achieving the maximum value.
\end{itemize}\par
3. Calculate $\mathbf{R}=(r_0,r_1,...,r_{N-1})$, where $r_{N-1}=\arg\max\limits_{0\leq i\leq L-1} t_{i,N-1}$ and $r_{k-1}=x_{r_{k},k}$ ($k$ is from $N-1$ to 0).\par
4. Calculate $\mathbf{F}=(f_0,f_1,...,f_{N-1})$, where $f_k=l_{r_k,k}$ ($k$ is from $0$ to $N-1$). Return $\mathbf{F}$.\par
The complexity of CLDA is $O(L^2N)$, where $L$ is the list size and $N$ is the length of codeword sentence.
An example of CLDA is illustrated as follows.\\
\begin{example}\label{exam_1}
Let the codeword transition probability matrix $\mathbf{P}$ be
\begin{equation}
\mathbf{P}=
\left(
  \begin{array}{cccc}
    1/3 & 2/3  & 0  \\
    2/9 & 5/9  & 2/9  \\
    0   & 2/3  & 1/3  \\
  \end{array}
\right)\nonumber
\end{equation}
Note that, $\mathbf{P}$ is from the context and has no relation with BSC.
Meanwhile, let the priori probability of each codeword be $\Pi=(1/5,3/5,1/5)^T$. There are $3$ successive lists with list size $3$, which is stored in $\mathbf{L}$.
\begin{equation}
\mathbf{L}=
\left(
  \begin{array}{ccc}
    0 & 0  & 0  \\
    1 & 1  & 1  \\
    2 & 2  & 2  \\
  \end{array}
\right)\nonumber
\end{equation}

1. Initialize the first column of $\mathbf{T}$ with $\Pi=(1/5,3/5,1/5)^T$ .\par
2. For each $2$-codeword sentence from List 0 to List 1, (\ref{equ_t}) and (\ref{equ_x}) are calculated as follows: $\pi_0*P_{00}=1/15$, $\pi_1*P_{10}=2/15$, $\pi_2*P_{20}=0.$
Then, record the biggest value $2/15$ in $\mathbf{T}$ and corresponding jumping point $x_{01}$:
$t_{01}=2/15,\ x_{01}=1.$

Similarly, calculate $\pi_0*P_{01}=2/15,\ \pi_1*P_{11}=1/3,\ \pi_2*P_{21}=2/15$. Then $t_{11}=1/3,\ x_{11}=1.$

And calculate $\pi_0*P_{02}=0,\ \pi_1*P_{12}=2/15,\ \pi_2*P_{22}=1/15$. Then $t_{21}=2/15,\ x_{21}=1.$

For each $3$-codeword sentence from List 0 to List 3, compute $t_{01}*P_{00}=2/45,\ t_{11}*P_{10}=2/27,\ t_{21}*P_{20}=0,$ and record the biggest value $2/27$  and corresponding index $1$. Then $t_{02}=2/27,\ x_{02}=1.$\par
Similarly, $t_{12}=5/27,\ x_{12}=1$, $t_{22}=2/27,\ x_{22}=1.$

3. We can find the most probable sentence using $\mathbf{T}$ and $\mathbf{A}$, as shown in Algorithm 1. $t_{12}$ is the maximum among $t_{02}$, $t_{12}$ and $t_{22}$. Then $r_2=1$ and $r_1=x_{12}=1$, $r_0=x_{11}=1$. Using $\mathbf{R}=(1,1,1)$, we get $\mathbf{F}=(l_{10},l_{11},l_{12})=(1,1,1)$ which is the final codeword sentence after context list decoding.
\end{example}

\section{Performance of CLDA}
\label{sec:5}

As the ARQ system transmits messages through BSC with crossover probability $p$, the probability of retransmission each time is
\begin{align*}
P_{retrans}=1-A^w_C(1-p,p),
\end{align*}
where we assume the sent codeword is $(0,0,...,0)_n$. In the following paragraphs,
$\alpha (\alpha\geq 0)$ denotes the number of retransmissions for receiving a legal codeword.

Let $\mathbf{s}=(s_0,s_1,...,s_{N-1})$ be the sent codeword sentence.
To guarantee that $\mathbf{s}$ can be selected from the $N$ lists in the decoding,
it is obvious that $\Pr\{\mathbf{s}\}$ should be the biggest among all the sentences in $L_0\times L_1\times...\times L_{N-1}$.
This constraint depends not only on the codeword sentence but also on the transition probability matrix $P$.

{\bf Assumption.} Assume that $\Pr\{\mathbf{s}\}$ is the biggest among all the codeword sentence in $\mathcal{L}(s_0)\times \mathcal{L}(s_1)\times...\times \mathcal{L}(s_{N-1})$. Then if received sentence $\mathbf{r}$ equals to $\mathbf{s}$, the CLDA selects $\mathbf{s}$ obviously. The following paragraphs try to consider the situation that $\mathbf{r}\neq\mathbf{s}$.

Assume $s_i\neq r_i$ for some $0\leq i\leq N-1$. If the list given by $s_i$ is the same with the list given by $r_i$, $\mathbf{s}$ will be selected from the lists correctly, otherwise $\mathbf{s}$ may be selected wrongly.

Let $S(s_i)$ denote the set of codewords which give the same list with $s_i$, namely
\begin{align}
S(s_i)=\{c\in C: \mathcal{L}(c)=\mathcal{L}(s_i)\}.\label{equ_S}
\end{align}

We use $E_c$ to denote the event that $\mathbf{s}$ is selected from the lists correctly. Meanwhile, let $E_c(i)(0\leq i\leq N-1)$ denote the event that $s_i$ is selected correctly from $L_i$. Then a lower bound of $\Pr\{E_c\}$ is as follows:
\begin{small}\begin{align}
&\Pr\{E_c\}=\Pr\{\bigcap_{i=0}^{N-1} E_c(i)\}=\prod_{i=0}^{N-1}\Pr\{E_c(i)\} \notag\\
&\geq \prod_{i=0}^{N-1}\Pr\{\mathcal{L}(r_i)=\mathcal{L}\{s_i\}\}\notag\\
&=\prod_{i=0}^{N-1}\Big(\sum_{\alpha=0}^\infty P_{retrans}^{\alpha}\Pr\{r_i \in S(s_i)\}\Big)\notag\\
&=\frac{1}{(1-P_{retrans})^N}\prod_{i=0}^{N-1}(P_c+(|S(s_i)|-1)p^d(1-p)^{n-d}).\label{equ_PrEc}
\end{align}\end{small}
Note that $\Pr\{r_i \in S(s_i)\}$ is calculated based on the definition of $\mathcal{L}(\cdot)$ and $S(\cdot)$, see (\ref{equ_L}) and (\ref{equ_S}).

\begin{figure*}[!t]
 \centering
  \subfigure[]
  {\includegraphics[width=0.32\textwidth]{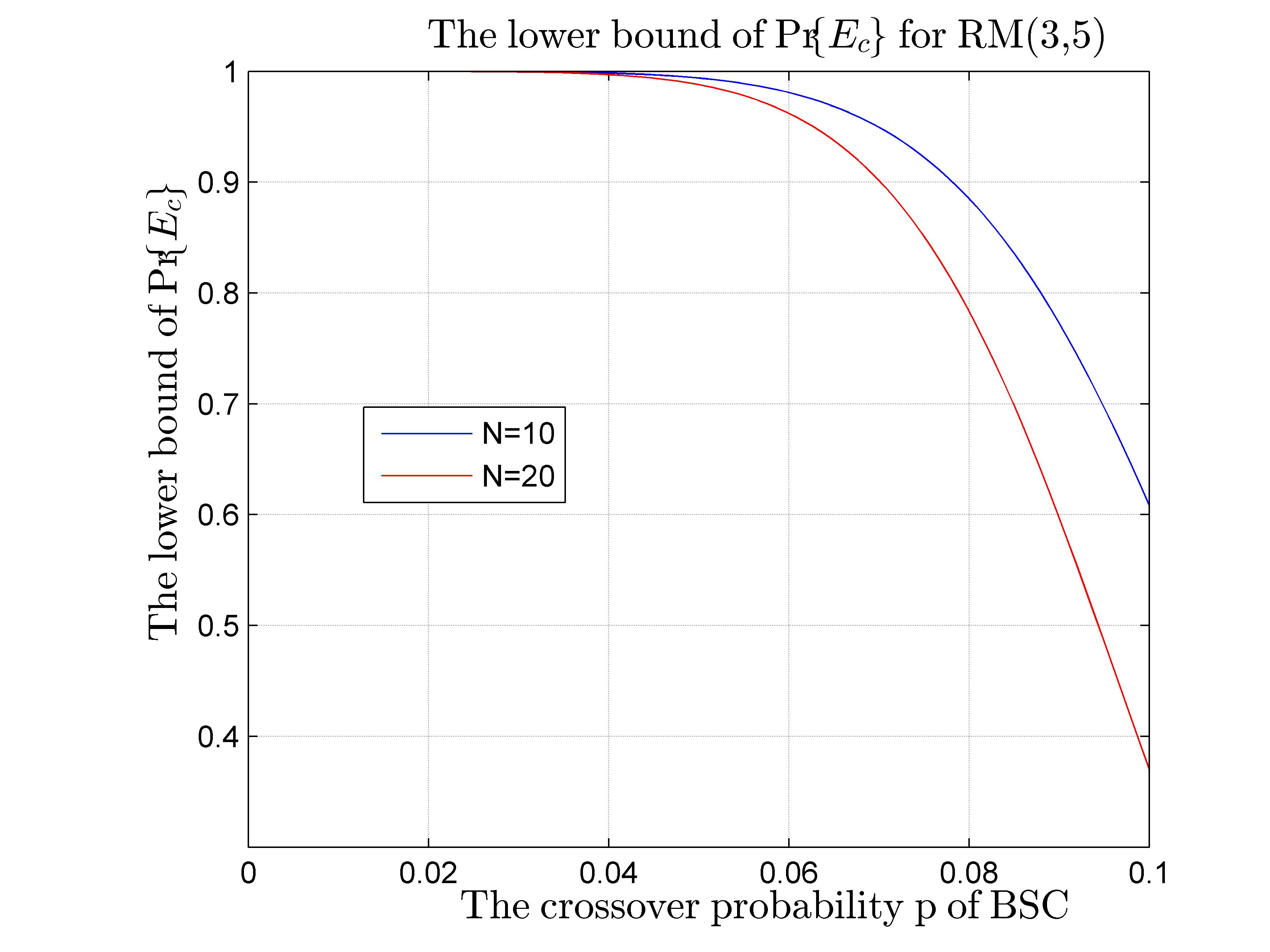}}
  \subfigure[]
  {\includegraphics[width=0.32\textwidth]{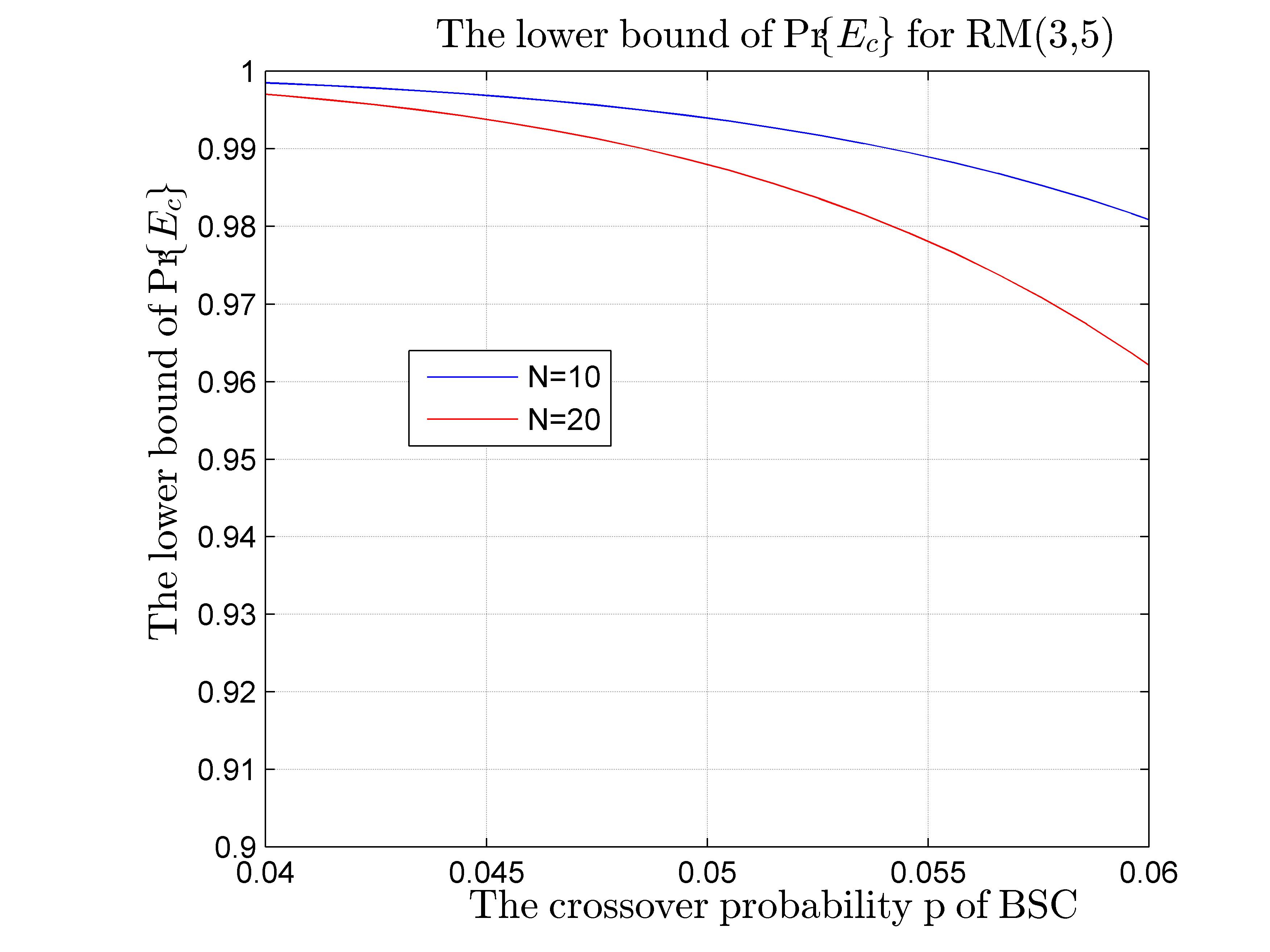}}
  \subfigure[]
  {\includegraphics[width=0.32\textwidth]{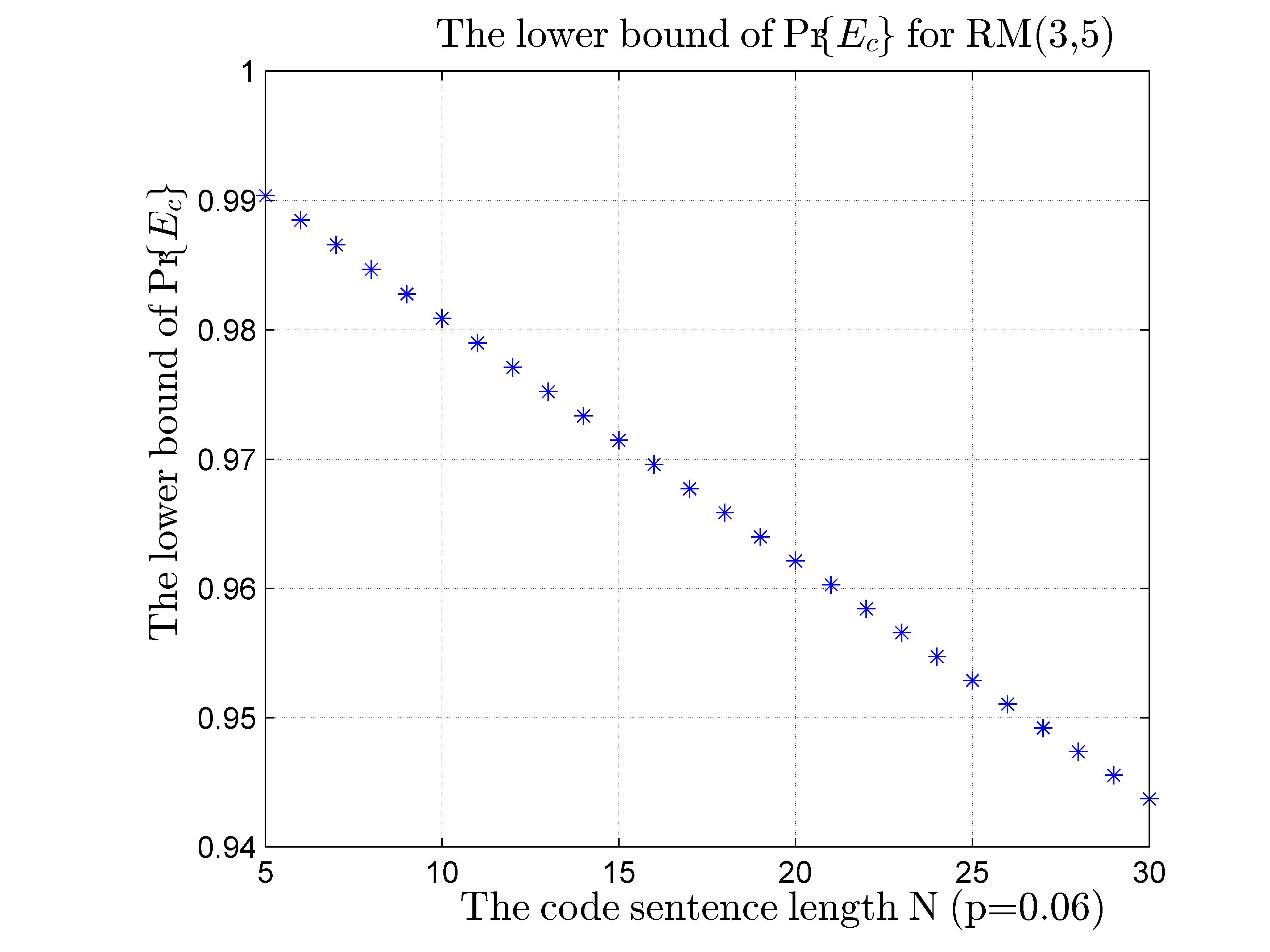}}
  \caption{The lower bound of correctly selected probability $\Pr\{E_c\}$}\label{fig_PrEc}
\end{figure*}

Numerical results for (\ref{equ_PrEc}) are shown in Fig. \ref{fig_PrEc}, where the lower bound of $P_{E_c}$ is evaluated with RM(3,5) code. In Fig. \ref{fig_PrEc} (a), the lower bound of $P_{E_c}$ decreases when the crossover probability $p$ of BSC varies from $0$ to $0.1$ for codeword sentence length $N=10,20$. It can be seen that the bound is better for small $p$, i.e. $p\in(0,0.06)$, which is partly show in Fig. \ref{fig_PrEc} (b). The lower bound is about $0.98$ for $p\in(0.04,0.06)$ which may be practical. Fig. \ref{fig_PrEc} (c) shows that the lower bound of $P_{E_c}$ decreases when the codeword sentence $N$ varying from $5$ to $30$ and $p=0.06$. For small $N$, $P_{E_c}$ may be relatively large, however, the context may not work well in CLDA. When $p$ and $P_{E_c}$ is fixed, the largest $N$ can be calculated out with (\ref{equ_PrEc}), which is meaningful for the selection of the proper codeword sentence length.

 When $p$ is fixed, the lower bound can be used to calculated the proper $N$ needed to achieve required correctly selected probability $P_{E_c}$.
%

In the following part, we try to analyse the average correctly selected probability denoted by $P_{average}$.

The probability of all the $M^N$ codeword sentences can be calculated out with $\mathbf{P}$. It is convenient to order all the codeword sentences by the probability from large to small.

Let $\mathbf{s}_i=(s_{i0},s_{i1},...,s_{i,N-1})(1\leq i\leq M^N)$ denote the ordered codeword sentences, where $\Pr\{\mathbf{s}_i\}$ is decreasing by $i$ from $1$ to $M^N$, namely, $\Pr\{\mathbf{s}_1\}$ is largest and $\Pr\{\mathbf{s}_{M^N}\}$ is the smallest.

For each sent codeword sentence $\mathbf{s}_i$, the probability that $\mathbf{s}_i$ is selected correctly varies with $\Pr\{\mathbf{s}_i\}$. Consider $\mathbf{s}_i$ and $\Pr\{\mathbf{s}_i\}$ for different $1\leq i\leq M^N$ respectively as follows:

1. $\mathbf{s}_1$ and $\Pr\{\mathbf{s}_1\}$:

   $\Pr\{\mathbf{s}_1\}$ is the largest among all the codeword sentences, so if $\mathbf{s}_1$ in the given lists, it will be selected correctly.
   \begin{align*}
   &\Pr\{\mathbf{s}_1 \text{ in the lists}|\mathbf{s}_1 \text{ is sent}\}\\
   &=\frac{1}{(1-P_{retrans})^N}(P_c+A_d\cdot p^d(1-p)^{n-d})^N
   \end{align*}

2. $\mathbf{s}_i$ and $\Pr\{\mathbf{s}_1\}$:

   If $\mathbf{s}_j(1\leq j\leq i)$ in the lists, $\mathbf{s}_i$ will not be selected out and errors happen. When $\mathbf{s}_i$ is sent, the given lists should contain $\mathbf{s}_i$ and not contain $\mathbf{s}_j(1\leq j\leq i)$.
   \begin{align*}
   &\Pr\{\mathbf{s}_i \text{ is selected correctly}|\mathbf{s}_i \text{ is sent}\}\notag\\
   &=\Pr\{\mathbf{r}\in \mathcal{L}(\mathbf{s}_i)\backslash\Big(\bigcup_{k=1}^{i-1} \mathcal{L(}\mathbf{s}_k)\Big) |\mathbf{s}_i \text{ is sent}\}\\
   &=\frac{1}{(1-P_{retrans})^N}\cdot\notag\\
   &\sum_{r\in\mathcal{L}(\mathbf{s}_i)\backslash (\bigcup_{k=1}^{i-1} \mathcal{L(}\mathbf{s}_k)) }\prod_{k=0}^{N-1}p^{d_H(r_k,s_{i,k})}(1-p)^{n-d_H(r_k,s_{i,k})},
   \end{align*}
   where
   $\mathcal{L}(\mathbf{s}_i)=\{\mathcal{L}(s_{i0})\times \mathcal{L}(s_{i1}) \times\mathcal{L}(s_{i,N-1})\}$
   and
   $\mathcal{L}(r_i)=\{c\in C: d_H(c,r_i)\leq d\} \text{, defined in (19)}. $

Then, the average correctly selected probability is
\begin{align}
&P_{average}\notag\\
&=\sum_{i=1}^{M^N}(\Pr\{\mathbf{s}_i \text{ is sent}\}\cdot \Pr\{\mathbf{s}_i \text{ is selected correctly}|\mathbf{s}_i \text{ is sent}\})\notag\\
&=\frac{1}{(1-P_{retrans})^N}\sum_{i=1}^{M^N}\Pr\{\mathbf{s}_i\}\cdot \notag\\
&\sum_{r\in\mathcal{L}(\mathbf{s}_i)\backslash (\bigcup_{k=1}^{i-1} \mathcal{L(}\mathbf{s}_k)) }\prod_{k=0}^{N-1}p^{d_H(r_k,s_{i,k})}(1-p)^{n-d_H(r_k,s_{i,k})}.\notag
\end{align}
To be convenient, we assume $\bigcup_{k=1}^{0} \mathcal{L(}\mathbf{s}_k)=\emptyset$.

$P_{average}$ can be calculated when the transition probability matrix $P$ is known. However, the calculation complexity is high. But when $P$ is sparse or with other special structures, the evaluation maybe useful. Besides, when $M$ and $N$ are relatively small, $P_{average}$ can be calculated and used to construct possible codes with better performance.

\section{Conclusion}
\label{Con}
In this paper, we use list decoding to correct undetected errors. The benefit is illustrated with numerical results of Hamming codes and Reed-Muller codes in ARQ systems, where the benefit proportion with list decoding can achieve about $40\%$ in some cases. We assume the code context model is a Markov chain and an algorithm based on context, named CLDA, is investigated to select the most probable codewords from lists. The performance of CLDA is analysed and a lower bound on the probability that the CLDA selects the correct codeword sentence is derived.

\end{document}